\documentclass[12pt]{article}
\usepackage{lineno}
\usepackage{graphicx}
\usepackage{amsmath}
\usepackage{hhline}
\usepackage{amssymb}
\usepackage{times}
\usepackage{cite}
\usepackage{array}
\usepackage{multirow}

 \usepackage{dcolumn}

\usepackage{color}
\definecolor{rltred}{rgb}{0.75,0,0}
\definecolor{rltgreen}{rgb}{0,0.5,0}
\definecolor{rltblue}{rgb}{0,0,0.75}

\usepackage{ifpdf}
\ifpdf
    \pdfoutput=1       
\fi

\ifpdf
\usepackage{thumbpdf}
\usepackage[
        colorlinks=true,
        urlcolor=rltblue,       
        filecolor=rltgreen,     
        linkcolor=rltred,       
        pdftitle={Example File for pdflatex},
        pdfauthor={H1},
        pdfsubject={Template file},
        pdfkeywords={High-Energy Physics, Particle Physics},
        pdfpagemode=None,
        bookmarksopen=true]{hyperref}
 
\fi

\newlength{\dinwidth}
\newlength{\dinmargin}
\begin{document}  

\newcolumntype{d}{D{.}{.}{-1}}
\newcolumntype{e}{D{/}{/}{-1}}
  
\newcommand{\pom}{{I\!\!P}}
\newcommand{\reg}{{I\!\!R}}
\newcommand{\slowpi}{\pi_{\mathit{slow}}}
\newcommand{\fiidiii}{F_2^{D(3)}}
\newcommand{\fiidiiiarg}{\fiidiii\,(\beta,\,Q^2,\,x)}
\newcommand{\n}{1.19\pm 0.06 (stat.) \pm0.07 (syst.)}
\newcommand{\nz}{1.30\pm 0.08 (stat.)^{+0.08}_{-0.14} (syst.)}
\newcommand{\fiidiiiful}{F_2^{D(4)}\,(\beta,\,Q^2,\,x,\,t)}
\newcommand{\fiipom}{\tilde F_2^D}
\newcommand{\ALPHA}{1.10\pm0.03 (stat.) \pm0.04 (syst.)}
\newcommand{\ALPHAZ}{1.15\pm0.04 (stat.)^{+0.04}_{-0.07} (syst.)}
\newcommand{\fiipomarg}{\fiipom\,(\beta,\,Q^2)}
\newcommand{\pomflux}{f_{\pom / p}}
\newcommand{\nxpom}{1.19\pm 0.06 (stat.) \pm0.07 (syst.)}
\newcommand {\gapprox}
   {\raisebox{-0.7ex}{$\stackrel {\textstyle>}{\sim}$}}
\newcommand {\lapprox}
   {\raisebox{-0.7ex}{$\stackrel {\textstyle<}{\sim}$}}
\def\gsim{\,\lower.25ex\hbox{$\scriptstyle\sim$}\kern-1.30ex%
\raise 0.55ex\hbox{$\scriptstyle >$}\,}
\def\lsim{\,\lower.25ex\hbox{$\scriptstyle\sim$}\kern-1.30ex%
\raise 0.55ex\hbox{$\scriptstyle <$}\,}
\newcommand{\pomfluxarg}{f_{\pom / p}\,(x_\pom)}
\newcommand{\dsf}{\mbox{$F_2^{D(3)}$}}
\newcommand{\dsfva}{\mbox{$F_2^{D(3)}(\beta,Q^2,x_{I\!\!P})$}}
\newcommand{\dsfvb}{\mbox{$F_2^{D(3)}(\beta,Q^2,x)$}}
\newcommand{\dsfpom}{$F_2^{I\!\!P}$}
\newcommand{\gap}{\stackrel{>}{\sim}}
\newcommand{\lap}{\stackrel{<}{\sim}}
\newcommand{\fem}{$F_2^{em}$}
\newcommand{\tsnmp}{$\tilde{\sigma}_{NC}(e^{\mp})$}
\newcommand{\tsnm}{$\tilde{\sigma}_{NC}(e^-)$}
\newcommand{\tsnp}{$\tilde{\sigma}_{NC}(e^+)$}
\newcommand{\st}{$\star$}
\newcommand{\sst}{$\star \star$}
\newcommand{\ssst}{$\star \star \star$}
\newcommand{\sssst}{$\star \star \star \star$}
\newcommand{\tw}{\theta_W}
\newcommand{\sw}{\sin{\theta_W}}
\newcommand{\cw}{\cos{\theta_W}}
\newcommand{\sww}{\sin^2{\theta_W}}
\newcommand{\cww}{\cos^2{\theta_W}}
\newcommand{\trm}{m_{\perp}}
\newcommand{\trp}{p_{\perp}}
\newcommand{\trmm}{m_{\perp}^2}
\newcommand{\trpp}{p_{\perp}^2}
\newcommand{\alp}{\alpha_s}

\newcommand{\alps}{\alpha_s}
\newcommand{\sqrts}{$\sqrt{s}$}
\newcommand{\LO}{$O(\alpha_s^0)$}
\newcommand{\Oa}{$O(\alpha_s)$}
\newcommand{\Oaa}{$O(\alpha_s^2)$}
\newcommand{\PT}{p_{\perp}}
\newcommand{\JPSI}{J/\psi}
\newcommand{\sh}{\hat{s}}
\newcommand{\uh}{\hat{u}}
\newcommand{\MP}{m_{J/\psi}}
\newcommand{\PO}{I\!\!P}
\newcommand{\xbj}{x}
\newcommand{\xpom}{x_{\PO}}
\newcommand{\ttbs}{\char'134}
\newcommand{\xpomlo}{3\times10^{-4}}  
\newcommand{\xpomup}{0.05}  
\newcommand{\dgr}{^\circ}
\newcommand{\pbarnt}{\,\mbox{{\rm pb$^{-1}$}}}
\newcommand{\gev}{\,\mbox{GeV}}
\newcommand{\WBoson}{\mbox{$W$}}
\newcommand{\fbarn}{\,\mbox{{\rm fb}}}
\newcommand{\fbarnt}{\,\mbox{{\rm fb$^{-1}$}}}
%
%
\newcommand{\qsq}{\ensuremath{Q^2} }
\newcommand{\gevsq}{\ensuremath{\mathrm{GeV}^2} }
\newcommand{\et}{\ensuremath{E_t^*} }
\newcommand{\rap}{\ensuremath{\eta^*} }
\newcommand{\gp}{\ensuremath{\gamma^*}p }
\newcommand{\dsiget}{\ensuremath{{\rm d}\sigma_{ep}/{\rm d}E_t^*} }
\newcommand{\dsigrap}{\ensuremath{{\rm d}\sigma_{ep}/{\rm d}\eta^*} }
\def\Journal#1#2#3#4{{#1} {\bf #2} (#3) #4}
\def\NCA{\em Nuovo Cimento}
\def\NIM{\em Nucl. Instrum. Methods}
\def\NIMA{{\em Nucl. Instrum. Methods} {\bf A}}
\def\NPB{{\em Nucl. Phys.}   {\bf B}}
\def\PLB{{\em Phys. Lett.}   {\bf B}}
\def\PRL{\em Phys. Rev. Lett.}
\def\PRD{{\em Phys. Rev.}    {\bf D}}
\def\ZPC{{\em Z. Phys.}      {\bf C}}
\def\EJC{{\em Eur. Phys. J.} {\bf C}}
\def\CPC{\em Comp. Phys. Commun.}

\newcommand{\ep}{ ep}
\newcommand{\ee}{ e^{+}e^{-}}
\newcommand{\xp}{ x_p}
\newcommand{\MeV}{\rm MeV}
\newcommand{\GeV}{\rm GeV}

\begin{titlepage}
\begin{flushleft}
{\tt DESY 09-084    \hfill    ISSN 0418-9833} \\
{\tt September  2009}                  \\
\end{flushleft}


\vspace{2cm}

\begin{center}
\begin{Large}

{\bf  Observation of the Hadronic Final State Charge Asymmetry in High $\bold{Q^{2}}$ Deep-Inelastic Scattering at HERA}

\vspace{2cm}
H1 Collaboration

\end{Large}
\end{center}

\vspace{2cm}

\begin{abstract}
A first measurement is presented of the charge asymmetry in the hadronic final state from the hard interaction in  deep-inelastic $\ep$ neutral current scattering at HERA. The measurement is performed in the range of negative squared four momentum transfer $100 < Q^{2} < 8,000~\GeV^{2}$. The difference between the event normalised distributions of the scaled momentum, $\xp$, for positively and negatively charged particles, measured in the current region of the Breit frame, is studied together with its evolution as a function of $Q$.
The results are compared to Monte Carlo models at the hadron and parton level.

\end{abstract}

\vspace{1.5cm}

\begin{center}
Accepted by Phys. Lett. B.
\end{center}

\end{titlepage}

%
%
%
\begin{flushleft}


F.D.~Aaron$^{5,49}$,           
M.~Aldaya~Martin$^{11}$,       
C.~Alexa$^{5}$,                
K.~Alimujiang$^{11}$,          
V.~Andreev$^{25}$,             
B.~Antunovic$^{11}$,           
A.~Asmone$^{33}$,              
S.~Backovic$^{30}$,            
A.~Baghdasaryan$^{38}$,        
E.~Barrelet$^{29}$
% ...

\bigskip{\it
% ...
$ ^{5}$ National Institute for Physics and Nuclear Engineering (NIPNE) ,
          Bucharest, Romania \\
% ...
$ ^{11}$ DESY, Hamburg, Germany \\
% ...
$ ^{25}$ Lebedev Physical Institute, Moscow, Russia$^{ e}$ \\
% ...
$ ^{29}$ LPNHE, Universit\'{e}s Paris VI and VII, IN2P3-CNRS,
          Paris, France \\
 $ ^{30}$ Faculty of Science, University of Montenegro,
          Podgorica, Montenegro$^{ e}$ \\
% ...
$ ^{33}$ Dipartimento di Fisica Universit\`a di Roma Tre
          and INFN Roma~3, Roma, Italy \\
% ...
$ ^{38}$ Yerevan Physics Institute, Yerevan, Armenia \\
% ...
$ ^{49}$ Also at Faculty of Physics, University of Bucharest,
          Bucharest, Romania \\
% ...
$ ^e$ Supported by the Deutsche Forschungsgemeinschaft \\
% ...
}

\end{flushleft}

\newpage

\section{Introduction}
\noindent
In lepton proton deep-inelastic scattering (DIS) at large Bjorken $x$ the contribution of $u$ valence quarks from the proton to the hard interaction dominates over that from the $d$ valence quarks due to their larger charge and greater abundance. Hence an asymmetry in the number of positively and negatively charged particles is observed in the final state~\cite{emc}. It has been demonstrated that the charge sign asymmetry of the hadronic final state in $pp$ collisions at RHIC~\cite{rhic} is sensitive to the valence quark distribution~\cite{akk2}. 

In a recent paper, H1 presented a study of the inclusive charged particle production in high $Q^{2}$ deep-inelastic scattering at HERA~\cite{h1162}. The measurement is performed in the current hemisphere of the Breit frame~\cite{breit}. In the na\"ive quark parton model (QPM) the momentum of the scattered parton in the Breit frame is $Q/2$, where $Q^{2}$ is the virtuality of the exchanged boson. The main observable is $\xp$, the charged particle momentum in the current region of the Breit frame scaled to $Q/2$. General agreement was observed between $\ep$, $\ee$ data and Monte Carlo predictions, broadly supporting the concept of quark fragmentation universality. Hadrons with small values of $\xp$ are predominately produced by fragmentation, while hadrons at large $\xp$ are more likely to contain a parton from the hard interaction. Therefore a study of the $\xp$ distribution separately for positively and negatively charged  particles should reveal information about the valence quarks and their fragmentation. 

The analysis presented here utilises the same data and methodology as in~\cite{h1162} but separates the positively and negatively charged particles into different distributions. In addition the charge asymmetry is studied. The results are compared with predictions from different fragmentation models implemented in Monte Carlo programs. 

\section{Experimental Method}
\label{Data Selection}

A full description of the H1 detector can be found elsewhere~\cite{h1detector} and only those components most relevant for this analysis are mentioned briefly here. The origin of the H1 coordinate system is the nominal $\ep$ interaction point, the direction of the proton beam defining the positive $z$--axis (forward region). 

The Liquid Argon (LAr) calorimeter measures the positions and energies of particles, including that of the scattered positron, over the polar angle range $4^\circ < \theta < 154^\circ$. The calorimeter consists of an electromagnetic section with lead absorbers and a hadronic section with steel absorbers. The energy resolution for electrons in the electromagnetic section is $\sigma(E)/E=11.5 \% / \sqrt{E} ~[\GeV]~\oplus 1\%$~\cite{Andrieu:1994yn}.

Charged particles are measured in the Central Tracking Detector (CTD) in the range \linebreak $20^\circ < \theta < 165^\circ$. The CTD comprises two large cylindrical Central Jet Chambers (CJCs) arranged concentrically around the beam-line, complemented by a silicon vertex detector~\cite{Pitzl:2000wz} covering the range $30^\circ < \theta < 150^\circ$, two $z$-drift chambers and two multiwire proportional chambers for triggering purposes, all within a solenoidal magnetic field of strength $1.16~\rm {T}$. The transverse momentum resolution is $\sigma(p_{T})/ p_{T} \simeq 0.006~p_{T}~[\GeV]~\oplus~0.02$~\cite{kleinwort2006} . In each event the tracks are used in a common fit procedure to determine the $\ep$ interaction vertex. 

The data used in this analysis correspond to an integrated luminosity of $44 ~\rm pb^{-1}$ and were taken by H1 in the year 2000 when protons with an energy of $920 ~\GeV$ collided with positrons with an energy of $27.5~\GeV$. The event and track selection follows that of~\cite{h1162}, here only the kinematic phase space is defined.  

The scattered positron is detected in the LAr calorimeter in the polar angular range $10^\circ < \theta_e < 150^\circ$ and with energy greater than $11~\GeV$. The negative squared four momentum transfer is required to be in the range $100 < Q^2 < 8\,000~\GeV^{2}$ and the inelasticity $y$, which is the fractional energy loss of the positron in the proton rest frame, to be in the range $0.05 < y < 0.6$. The polar scattering angle for a massless parton, calculated from the positron kinematics in the quark-parton model (QPM) approximation\footnote{In the approximation of a massless scattered quark, the polar scattering angle is given by $\theta_{q,lab}=~\cos^{-1}((xs(xs-Q^{2})-4E^{2}Q^{2})/(xs(xs-Q^{2})+4E^{2}Q^{2}))$, where $E$ is the incoming positron beam energy, $s$ is the $\ep$ centre of mass energy squared and Bjorken $x$ is the fraction of the proton momentum carried by the struck quark in the QPM.\color{black}}, is required to be in the range $30^\circ < \theta_{q,lab} < 150^\circ$. This ensures that the current region of the Breit frame remains in the central region of the detector where there is high acceptance and track reconstruction efficiency. It should be noted that the kinematic phase space is defined solely from the scattered electron and can be applied in a simple way to theoretical models. The full event selection outlined above results in a data sample of about 60,000 events.

\section{Observables}

The current hemisphere of the Breit frame of reference provides a kinematic region where the properties of the scattered quark can be studied with a well defined and relatively clean separation from the proton remnants. In the Breit frame the virtual space-like photon has momentum $Q$ but no energy. The photon direction defines the negative $z'$--axis and the current hemisphere. 
Within the QPM the photon collides head on with a massless quark of longitudinal momentum $Q/2$. The struck quark thus scatters with an equal but opposite momentum into the current hemisphere while the proton remnants go into the opposite (target) hemisphere. The energy scale is set by the virtual photon at $Q/2$.
The boost to the Breit frame is defined using kinematics calculated from the scattered positron and is thus not biased by mismeasurements of the hadronic final state. Hadrons emerging from the interaction with negative longitudinal momenta in this frame are assigned to the current region and associated with the struck quark.

The scaled momentum variable, $\xp$, is defined to be $p_{h}/(Q/2)$ where $p_{h}$ is the momentum of a charged track in the current region of the Breit frame. The inclusive, event normalised, charged particle scaled momentum distribution, $D(\xp,Q)$, is calculated as $\frac{1}{N} \frac {dn} {d\xp}$, where in each $Q$ range, $N$ is the number of selected events and $dn$ is the total number of charged tracks with scaled momentum $\xp$ in the interval $d\xp$. The distribution $D(\xp,Q)$ has been measured previously \cite{h1162}.

In the present study, the scaled momentum distribution is defined separately for positively charged particles, $D^{+}(\xp,Q)$, and for negatively charged particles, $D^{-}(\xp,Q)$. The charge asymmetry is defined as $A(\xp,Q)=(D^{+}(\xp,Q)-D^{-}(\xp,Q)) / D(\xp,Q)$, where the relation $D(\xp,Q)=D^{+}(\xp,Q)+D^{-}(\xp,Q)$ holds. Systematic errors mostly cancel in the charge asymmetry. 

\section{Phenomenology}
\label{sec:Phenomenology}

A more complete discussion on the different models of the parton cascade and hadronisation processes can be found in~\cite{h1162}. 

The data are compared  to predictions of the Parton Shower model (PS)~\cite{ps}, as implemented in the RAPGAP~\cite{rapgap} Monte Carlo program and to predictions of the Colour Dipole Model (CDM)~\cite{cdm} both matched to $O(\alpha_{s})$ matrix elements. ARIADNE~\cite{ariadne} provides an implementation of CDM and is used in the DJANGO~\cite{django}  Monte Carlo program. Both the PS and CDM predictions use the Lund string model for hadronisation~\cite{lund}. The HERWIG Monte Carlo~\cite{herwig} program uses the parton shower model to describe the fragmentation process but incorporates the cluster model of hadronisation~\cite{cluster}.

In the Soft Colour Interaction model (SCI)~\cite{sci} soft gluons are exchanged between the partons produced in the parton shower and the proton remnant. A refined version of the model uses a generalised area law (GAL)~\cite{gal} for the colour rearrangement probability. Predictions for SCI and GAL models are obtained using the LO generator programs LEPTO~\cite{lepto} for DIS. Higher order QCD effects are simulated using parton showers. Hadronisation is simulated using the Lund string model. The SCI and GAL models produce similar predictions for the charge asymmetry, so only GAL model predictions are given in this paper. 

It is possible to turn off the hadronisation and compare data with parton level predictions using the assumption of local parton hadron duality. This has been done with the CDM predictions where a quark, with fractional charge, is taken as equivalent to a charged hadron of unit charge. The predictions are made after the main parton cascade has taken place and the gluons are ignored.

The CTEQ5L \cite{cteq5} parton density function (PDF) is used for all model predictions. Other PDFs~\cite{mrst2004, h1pdf2000} lead to charge asymmetries in agreement with the prediction based on CTEQ5L to within $\pm~0.01$.

\section{Data Correction}

The data are corrected for detector acceptance, efficiency and resolution effects using Monte Carlo event samples generated with the RAPGAP and DJANGO programs. All generated events are passed through the GEANT~\cite{geant} based simulation of the H1 apparatus and are reconstructed and analysed using the same programs as used for the data. These Monte Carlo event samples give a good description of the data. The residual contribution of charged particles from the weak decay of neutral particles (e.g. $K^{0}$ and $\Lambda$'s) is about $8\%$ and is subtracted from the data as part of the correction procedure. The effects of QED radiation are corrected for using the HERACLES~\cite{django} program incorporated within the above Monte Carlos. The total correction factor is calculated using DJANGO from the ratio of the number of entries in each bin at hadron level to that at detector level. The bin sizes are chosen to give high acceptance and purity\footnote{The acceptance (purity) is defined as the ratio of the number of charged hadrons generated and reconstructed to the total number of charged hadrons generated (reconstructed) in that bin.}, typically above $60\%$ with a minimum of $40\%$. The total correction factor applied to the uncorrected data points is typically  $1.0 - 1.2$ for $D^{\pm}(x_{p},Q)$. In general the uncertainty in the boost to the Breit frame dominates the resolution in $\xp$. The correction associated with the tracking dominates in the highest $Q^2$ region where there is a somewhat reduced acceptance for the current region of the Breit frame within the CTD. 

In the measurement of the charge asymmetry $A(\xp,Q)$ contributions to the correction factor such as QED corrections, efficiencies and acceptance mostly cancel and as a result the correction factor is consistent with 1.0. 

The data correction method was cross checked using a matrix migration unfolding procedure, which was found to be in agreement.

\section{Systematic Uncertainties}

The positron energy scale uncertainty is $0.7 - 3~\% $  depending on the position of the detected positron in the LAr calorimeter. This uncertainty affects both the phase space and boost calculation. The resulting uncertainty on $D^{\pm}(x_{p},Q)$ is independent of  $Q$ but varies with $x_{p}$ from $0.5\%$ ($x_{p}\sim 0.1$) to $11\%$ ($x_{p} \sim1.0$). The precision in the reconstruction of the scattered positron direction leads to a systematic error of about $1\%$. The hadronic energy scale uncertainty is $4\%$ and this leads to an error of less than $1\%$ on $D^{\pm}(\xp,Q)$. The uncertainty in the correction factor arising from using different Monte Carlo models in the correction procedure, taken as the full difference between correcting the data with RAPGAP or DJANGO, results in a typical error of $1.5\% $ on $D^{\pm}(\xp,Q)$. In the defined kinematic region, errors arising from non $\ep$ background are negligible. The systematic error associated with the track reconstruction (e.g. track reconstruction efficiency, vertex reconstruction efficiency, weak decays and nuclear interaction uncertainties) is estimated to be $2.5\%$ for  $D^{\pm}(x_{p},Q)$. All sources of error are treated as uncorrelated, apart from the positron energy scale uncertainty which is treated as fully correlated between bins. 

For the charge asymmetry measurements the effects of the systematic errors mostly cancel and the only significant contribution is from the track reconstruction uncertainty. The track reconstruction efficiency for positive and negative tracks is conservatively considered as fully anti correlated in the estimation of the effect on the charge asymmetry. Secondary interactions of tracks in the material in front of the trackers may also lead to charge asymmetries. At large track momentum a possible charge bias in the track reconstruction is studied using the track of the scattered lepton in $e^{+}p$ and $e^{-}p$ data (in the year 1998-1999 HERA operated with an electron beam). These sources of systematic error result in an uncertainty of $2.5\%$ on the charge asymmetry.

\section{Results}

\begin{figure}[h] 
  \begin{center}
    \includegraphics[width=15cm]{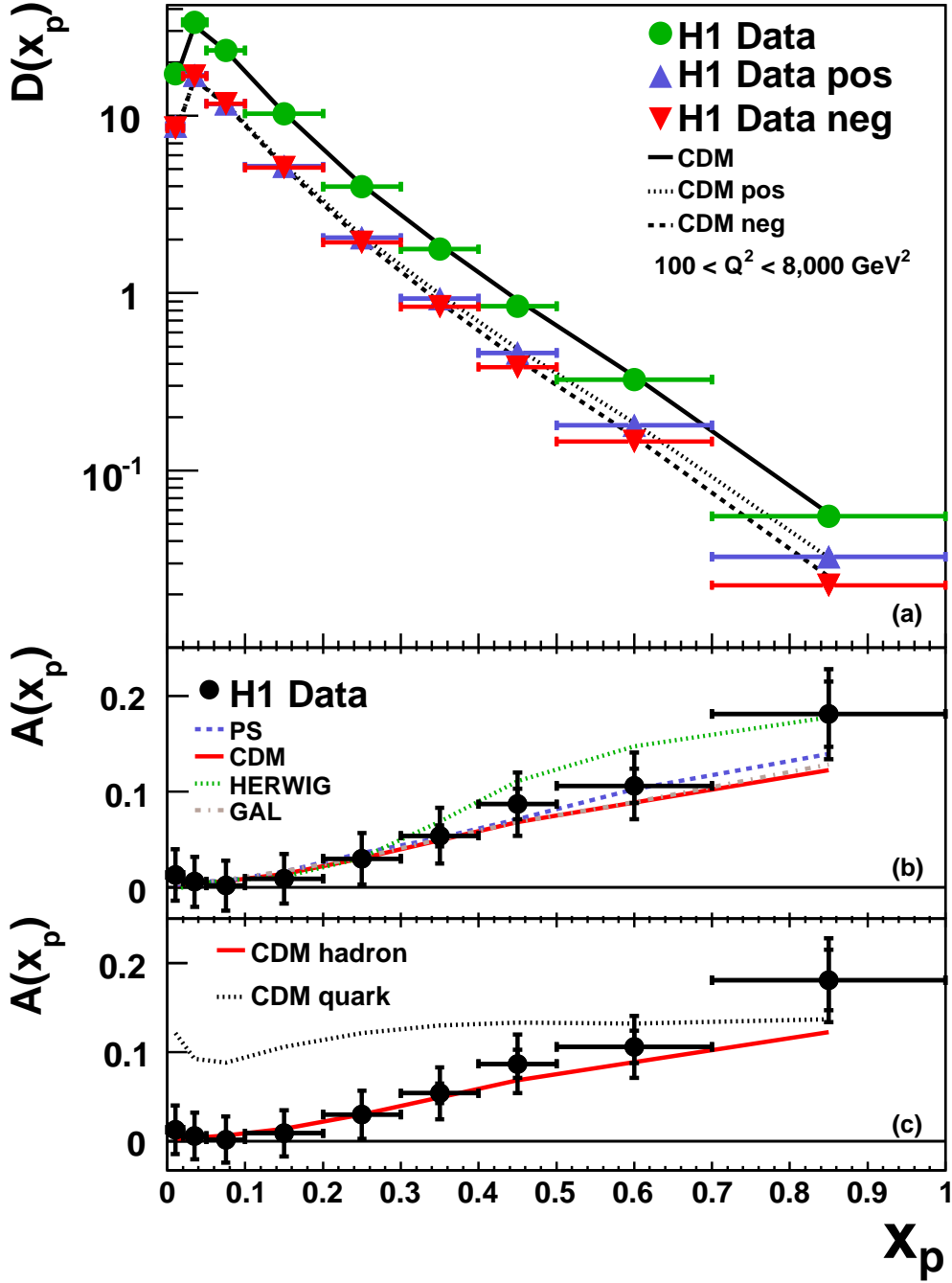}
  \end{center}
  \caption{\label{fig:summarydis}(a) The measured normalised distributions of the scaled momentum, $D(\xp)$, for all charged particles and for positively (pos), $D^{+}(\xp)$, and negatively (neg), $D^{-}(\xp)$, charged particles, and (b, c) the charge asymmetry, $A(\xp)$, as a function of $x_{p}$. The error bars include statistical (inner), and statistical plus systematic errors added in quadrature (outer). The data are compared to predictions from different models of the parton cascade and hadronisation processes implemented in leading order matrix element Monte Carlo programs and to the parton level before hadronisation.}
\end{figure}

The scaled momentum distribution for all charged particles, and for positively, and negatively charged particles separately, is shown in figure~\ref{fig:summarydis}a)  and table~\ref{table:xpsplit}. There are significantly more particles produced at low $x_{p}$ than at high $x_{p}$, in agreement with predictions. The scaled momentum distribution for positively and negatively charged particles are very similar at low $x_{p}$ but at high $x_{p}$ there is a clear excess of positively charged particles. 

The charge asymmetry can be as large as $0.18$ as shown in figure~\ref{fig:summarydis}b)  and table~\ref{table:asymmetry}. The scaled momentum distribution and its asymmetry is described by the Monte Carlo models. Models incorporating string hadronisation (PS, CDM, GAL) produce a smaller charge asymmetry at high $x_{p}$ than that produced by the cluster hadronisation model (HERWIG). Similar differences between the models at high $x_{p}$ were observed in~\cite{h1162}.

In figure~\ref{fig:summarydis}c) the charge asymmetry is compared to predictions from CDM before and after hadronisation. It is observed that at high $x_{p}$ the hadron and quark levels are in good agreement and both agree with the data. However, as $x_{p}$ gets smaller a large difference develops with the quark level asymmetry prediction constant at about $0.12$, while the hadron level, and the data, fall to zero. This is consistent with the expectation that the hadrons at low $x_{p}$ are dominantly produced  by fragmentation while hadrons at high $x_{p}$ retain the memory of the charge of the scattered quark from the hard interaction. It should be noted that sea quarks and gluons will produce, on average, charge symmetric hadronic final states, reducing the charge asymmetry expected from valence quarks alone.

\begin{figure}[h] 
  \begin{center}
    \includegraphics[width=15cm]{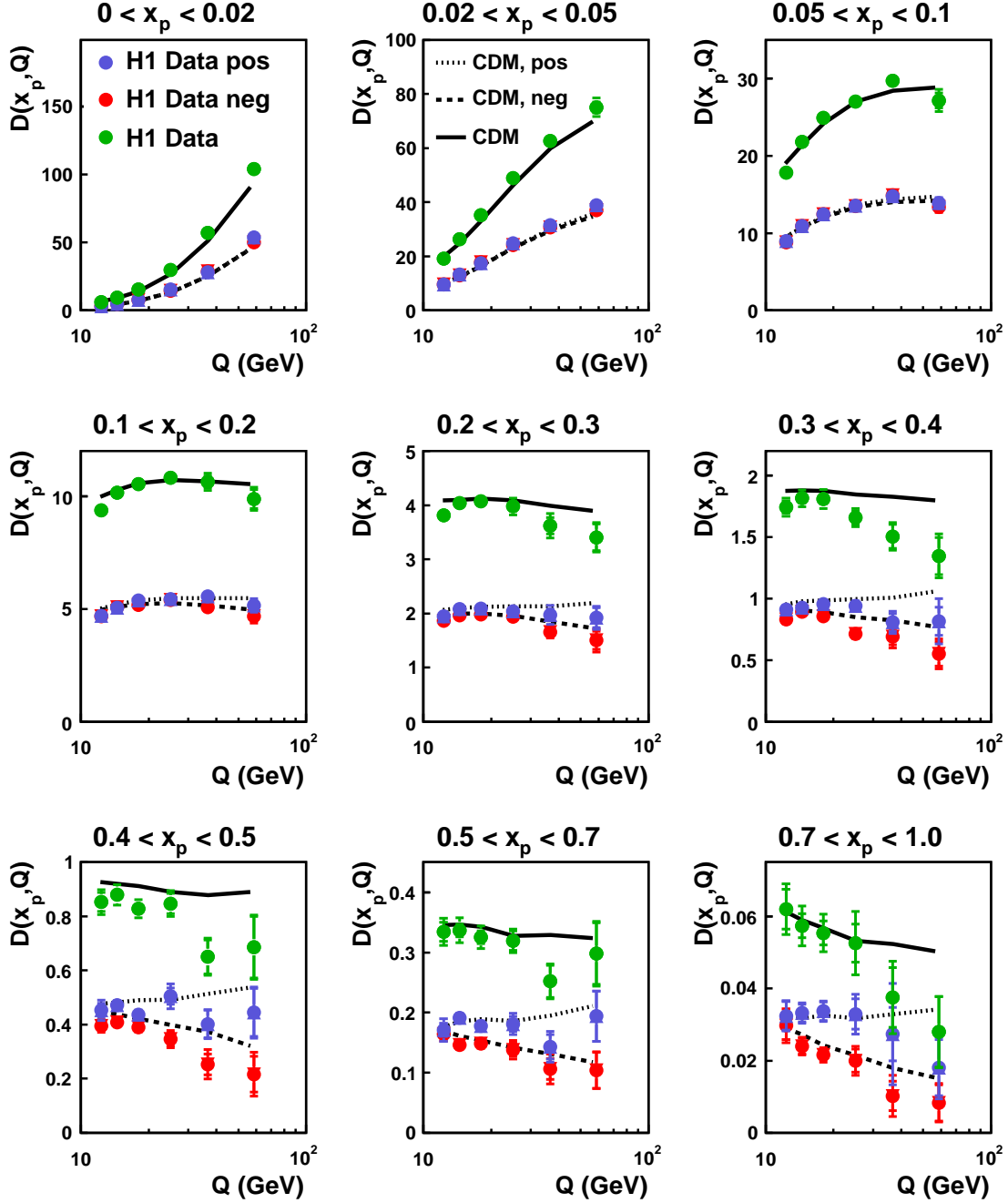}
  \end{center}
  \caption{ \label{fig:fragmentationdis} The measured normalised distributions of the scaled momentum, $D(\xp, Q)$, for all charged particles and for positively (pos), $D^{+}(\xp,Q)$, and negatively (neg), $D^{-}(\xp,Q)$, charged particles separately, as a function of $Q$ for nine different $\xp$ regions. The error bars include statistical (inner), and statistical plus systematic errors added in quadrature (outer). The data are displayed at the average value of $Q$. The data are compared to predictions from the CDM Monte Carlo program.}
\end{figure}

\begin{figure}[h] 
  \begin{center}
    \includegraphics[width=15cm]{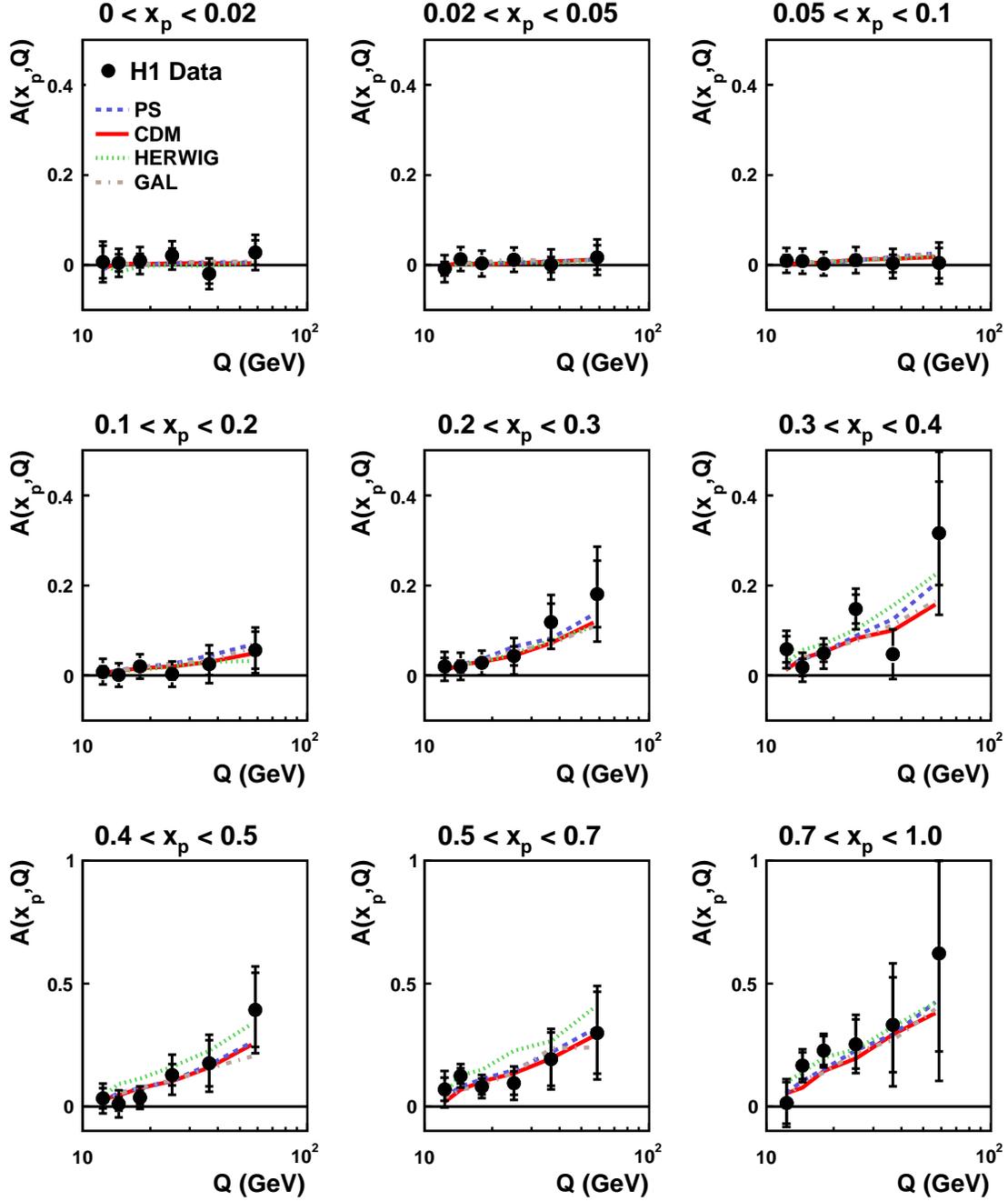}
  \end{center}
  \caption{ \label{fig:asymmetrydis} The charge asymmetry, $A(\xp,Q)$, as a function of $Q$ for nine different $\xp$ regions. The error bars include statistical (inner), and statistical plus systematic errors added in quadrature (outer). The data are displayed at the average value of $Q$. The data are compared to predictions from different models of the parton cascade and hadronisation processes implemented in leading order matrix element Monte Carlo programs.}
\end{figure}

The analysis intervals in $Q^{2}$ and the average values of $Q$ and Bjorken $x$ are shown in table~\ref{table:qandx}. Figures~\ref{fig:fragmentationdis} and~\ref{fig:asymmetrydis} (tables~\ref{table:xpqsplit} and \ref{table:asymmetryq}) show the scaled momentum distribution and charge asymmetry as a function of $Q$ in different $x_{p}$ intervals. The charge asymmetry observed at large $\xp$ evolves to larger values as $Q$ increases. The largest asymmetries of about $0.4$ are obtained in the highest $Q$ and highest $x_{p}$ intervals. It should be noted that higher average $Q$ corresponds to higher average Bjorken $x$ (table~\ref{table:qandx}) and hence the highest $Q$ intervals are most sensitive to the valence quark distribution. The scaled momentum distributions ($D^{\pm}(\xp,Q)$) are broadly predicted by the Monte Carlo predictions but tend to undershoot at large $Q$, similar to their sum~\cite{h1162}. The charge asymmetry is well described by the Monte Carlo.

\section{Conclusions}

The first measurement of the charge asymmetry of the hadronic final state at HERA is presented. The charge asymmetry is found to be dependent on the scaled momentum $x_{p}$ with a larger asymmetry for large $x_{p}$. The observed charge asymmetry at large $\xp$ is found to increase with the scale $Q$ corresponding at HERA to an enhancement at large Bjorken $x$. The results are consistent with the expectation that at high $\xp$ the asymmetry is directly related to the valence quark content of the proton. The observed charge asymmetry is reproduced by various models. The data are expected to provide useful information for the extraction of fragmentation functions and additional constraints on the valence quark distribution of the proton.

\section*{Acknowledgements}

We are grateful to the HERA machine group whose outstanding
efforts have made this experiment possible. 
We thank
the engineers and technicians for their work in constructing and
maintaining the H1 detector, our funding agencies for 
financial support, the
DESY technical staff for continual assistance
and the DESY directorate for support and for the
hospitality which they extend to the non DESY 
members of the collaboration.


\newpage

\begin{table}[h]
\renewcommand{\arraystretch}{1.2}
\begin{center}
\begin{tabular}{|c||d|d|d|e|}
\hline
\multicolumn{1}{|c||}{$x_{p}$} & 
\multicolumn{1}{c|}{$D^{(\pm)}(x_{p})$} & 
\multicolumn{1}{c|}{$\delta_{stat}[\%]$} & 
\multicolumn{1}{c|}{$\delta_{tot}[\%]$} & 
\multicolumn{1}{c|}{$\delta_{scale}[\%]$} \\
\hline
& \multicolumn{4}{c|}{sum} \\
\hline
$~0.0 < x_{p} < 0.02$ & 17.26 & 0.8 & 2.8 & 0.7 / 0.5 \\
\hline
$0.02 < x_{p} < 0.05$ & 33.55 & 0.5 & 2.6 & 1.1 / 1.2 \\
\hline
$0.05 < x_{p} < 0.1~$ & 23.35 & 0.4 & 2.6 & 1.3 / 1.3 \\
\hline
$0.1 < x_{p} < 0.2$ & 10.27 & 0.5 & 2.6 & 1.4 / 1.6 \\
\hline
$0.2 < x_{p} < 0.3$ & 3.98 & 0.7 & 2.9 & 2.0 / 1.9 \\
\hline
$0.3 < x_{p} < 0.4$ & 1.766 & 1.1 & 3.2 & 2.6 / 2.9 \\
\hline
$0.4 < x_{p} < 0.5$ & 0.843 & 1.6 & 3.6 & 3.8 / 3.4 \\
\hline
$0.5 < x_{p} < 0.7$ & 0.326 & 1.8 & 3.6 & 5.0 / 5.1 \\
\hline
$0.7 < x_{p} < 1.0$ & 0.0553 & 3.4 & 5.5 & 9.0 / 8.7 \\
\hline
& \multicolumn{4}{c|}{positive} \\
\hline
$~0.0 < x_{p} < 0.02$ & 8.74 & 1.1& 3.0 &  0.7 /  0.4\\
\hline
$0.02 < x_{p} < 0.05$ & 16.87 & 0.7 & 2.7 &  1.1 / 1.1\\
\hline
$0.05 < x_{p} < 0.1~$ & 11.70 & 0.7 & 2.7 & 1.3 / 1.3\\
\hline
$0.1 < x_{p} < 0.2$ & 5.18 & 0.7 & 2.7 &  1.3 / 1.4\\
\hline
$0.2 < x_{p} < 0.3$ & 2.05 & 1.0 & 2.9 & 1.8 / 1.6\\
\hline
$0.3 < x_{p} < 0.4$ & 0.930 & 1.5 & 3.4 & 2.7 / 3.1\\
\hline
$0.4 < x_{p} < 0.5$ & 0.460 & 2.2 & 3.9 & 3.9 / 3.3\\
\hline
$0.5 < x_{p} < 0.7$ & 0.180 & 2.5 & 4.0 & 4.1 / 5.0 \\
\hline
$0.7 < x_{p} < 1.0$ & 0.0327 & 4.6 & 6.3 &  9.0 / 9.0\\
\hline
& \multicolumn{4}{c|}{negative} \\
\hline
$~0.0 < x_{p} < 0.02$ & 8.52 & 1.2 & 3.0 & 0.7 / 0.5\\
\hline
$0.02 < x_{p} < 0.05$ &  16.68 & 0.7 & 2.7 &  1.2 / 1.3\\
\hline
$0.05 < x_{p} < 0.1~$ &  11.66 & 0.6 & 2.7 & 1.3 / 1.3\\
\hline
$0.1 < x_{p} < 0.2$ &  5.09 & 0.7 & 2.7 & 1.4 / 1.7\\
\hline
$0.2 < x_{p} < 0.3$ &  1.93 & 1.1  & 2.9 & 2.2 / 2.2\\
\hline
$0.3 < x_{p} < 0.4$ &  0.836 & 1.6 & 3.4 & 2.6 / 2.7\\
\hline
$0.4 < x_{p} < 0.5$ &  0.383 & 2.4 & 4.0 & 3.7 / 3.4\\
\hline
$0.5 < x_{p} < 0.7$ &  0.146  & 2.7 & 4.1 & 5.6 / 5.1\\
\hline
$0.7 < x_{p} < 1.0$ & 0.0226 & 5.4 & 6.9 & 9.0 / 11.0\\
\hline
\end{tabular}
\end{center}
\caption{\label{table:xpsplit} The measured normalised distribution of the scaled momentum for all charged particles, $D(\xp)$, and for positively, $D^{+}(x_{p})$, and negatively, $D^{-}(x_{p})$, charged particles, for different $x_{p}$ intervals, shown with the statistical error ($\delta_{stat}$), the total error including statistical and systematic errors added in quadrature ($\delta_{tot}$ ), and the correlated error coming from the electron energy scale uncertainty ($\delta_{scale}$) which is shown as two errors ($+/-$) and not included in the total error. The average $Q$ value for the data is $19.5~\GeV$.}
\end{table}%

\begin{table}[h]
\renewcommand{\arraystretch}{1.2}
\begin{center}
\begin{tabular}{|c||d|d|d|}
\hline
\multicolumn{1}{|c||}{$x_{p}$} & 
\multicolumn{1}{c|}{$A(x_{p})$} &
\multicolumn{1}{c|}{$\delta_{stat}$} & 
\multicolumn{1}{c|}{$\delta_{tot}$} \\
\hline
\hline
$~0.0 < x_{p} < 0.02$ & 0.013 & 0.008 & 0.027 \\
\hline
$0.02 < x_{p} < 0.05$ & 0.006 & 0.005 & 0.026 \\
\hline
$0.05 < x_{p} < 0.1~$ & 0.002 & 0.004 & 0.026 \\
\hline
$0.1 < x_{p} < 0.2$ & 0.009 & 0.005 & 0.026 \\
\hline
$0.2 < x_{p} < 0.3$ & 0.030 & 0.007 & 0.027 \\
\hline
$0.3 < x_{p} < 0.4$ & 0.054 & 0.011 & 0.029 \\
\hline
$0.4 < x_{p} < 0.5$ & 0.087 & 0.016 & 0.033 \\
\hline
$0.5 < x_{p} < 0.7$ & 0.106 & 0.018 & 0.035 \\
\hline
$0.7 < x_{p} < 1.0$ & 0.181  & 0.034 & 0.047 \\
\hline
\end{tabular}
\end{center}
\caption{\label{table:asymmetry} The charge asymmetry, $A(x_{p})$, for different $x_{p}$ intervals, shown with the statistical error ($\delta_{stat}$), the total error including statistical and systematic errors added in quadrature ($\delta_{tot}$). The average $Q$ value for the data is $19.5~\GeV$.}
\end{table}%

\begin{table}[h]
\renewcommand{\arraystretch}{1.2}
\begin{center}
\begin{tabular}{|c||d|d|d|d|}
\hline
\multicolumn{1}{|c||}{$Q^{2}~[\GeV^{2}]$} & 
\multicolumn{1}{c|}{$\langle Q \rangle~[\GeV]$} & 
\multicolumn{1}{c|}{$\delta Q ~[\GeV]$} & 
\multicolumn{1}{c|}{$\langle x \rangle $} & 
\multicolumn{1}{c|}{$\delta x$} \\
\hline
\hline
$100 < Q^{2} < 175$ & 12.3 & 0.1 &  0.00370 & 0.00004 \\
\hline
$175 < Q^{2} < 250$ & 14.5 & 0.1 & 0.00952 &  0.00007 \\
\hline
$250 < Q^{2} < 450$ & 18.0 & 0.1 &  0.1559 &  0.0001 \\
\hline
$~450 < Q^{2} < 1000$ & 25.0 & 0.3 &  0.0254 &  0.0003 \\
\hline
$1000 < Q^{2} < 2000$ & 36.6 & 0.8 &  0.044 & 0.001 \\
\hline
$2000 < Q^{2} < 8000$ & 58.5 & 2.1 &  0.087 &  0.003 \\
\hline
\end{tabular}
\end{center}
\caption {\label{table:qandx} Average $Q$ and $x$ values and their statistical errors for the selected events in the $Q^{2}$ intervals used in this analysis. The average $Q$ value for all data is $19.5~\GeV$.}
\end{table}%

\newpage

\begin{table}[htdp]
\renewcommand{\arraystretch}{1.2}
\begin{tiny}
\begin{center}
\begin{tabular}{|c||d|d|d|e||d|d|d|e|}
\hline
& \multicolumn{4}{c||}{positive} & \multicolumn{4}{c|}{negative} \\
\hline
\multicolumn{1}{|c||}{$Q^{2}~[\GeV^{2}]$} & 
\multicolumn{1}{c|}{$D^{+}(\xp,Q)$} & 
\multicolumn{1}{c|}{$\delta_{stat}$~[\%] } & 
\multicolumn{1}{c|}{$\delta_{tot}$~[\%]}  & 
\multicolumn{1}{c||}{$\delta_{scale}$~[\%]} & 
\multicolumn{1}{c|}{$D^{-}(\xp,Q)$} & 
\multicolumn{1}{c|}{$\delta_{stat}$~[\%]}  & 
\multicolumn{1}{c|}{$\delta_{tot}$~[\%]}  & 
\multicolumn{1}{c|}{$\delta_{scale}$~[\%]} \\
\hline
\hline
&\multicolumn{8}{c|}{ \rule{0mm}{2.5 mm} \raisebox{0.3mm}{$0.0 < x_{p} < 0.02$}} \\
\hline
$100 < Q^{2} < 175$ & 3.14 & 5.2 & 6.5 &  1.6 / 0.3 & 3.08 & 5.3 & 6.6 & 1.5 / 0.8 \\
\hline
$175 < Q^{2} < 250$ & 4.79 & 2.7 & 3.9 & 0.8 / 0.6 & 4.75 & 2.7 & 4.0 & 0.7 / 0.7 \\
\hline
$250 < Q^{2} < 450$ & 7.87 & 2.1 & 3.5 & 1.0 / 0.3 & 7.71 & 2.1 & 3.4 & 0.3 / 0.3 \\
\hline
$~450 < Q^{2} < 1000$ & 15.39 & 2.3 & 3.6 & 0.8 / 0.4 & 14.56 & 2.3 & 3.5 & 0.3 / 0.8 \\
\hline
$1000 < Q^{2} < 2000$ & 28.00 & 3.2 & 4.4 & 0.2 / 1.0 & 28.62 & 3.2 & 4.1 & 0.4 / 0.8\\
\hline
$2000 < Q^{2} < 8000$ & 53.58 & 3.8 & 4.6 & 0.7 / 0.7 & 50.23 & 3.9 & 4.8 & 0.8 / 1.3 \\
\hline
&\multicolumn{8}{c|}{ \rule{0mm}{2.5 mm} \raisebox{0.3mm}{$0.02 < x_{p} < 0.05$}}\\
\hline
$100 < Q^{2} < 175$ & 9.52 & 2.2 & 4.0 & 1.4 / 0.8 & 9.54 & 2.2 & 3.4 & 1.2 / 1.4 \\
\hline
$175 < Q^{2} < 250$ &13.29 & 1.3 & 2.9 & 1.0 / 1.2 & 12.97 & 1.3 & 2.9 & 1.2 / 1.7 \\
\hline
$250 < Q^{2} < 450$ &17.52 & 1.1 & 2.9 & 1.0 / 1.2 & 17.72 & 1.1 & 3.0 & 1.0 / 1.1 \\
\hline
$~450 < Q^{2} < 1000$ & 24.84 & 1.5 & 3.0 & 0.8 / 1.4 & 24.14 & 1.5 & 3.0 & 1.2 / 1.4\\
\hline
$1000 < Q^{2} < 2000$ & 31.55 & 2.5 & 3.6 & 0.5 / 1.0 & 30.64 & 2.5 & 4.5 & 0.8 / 1.1 \\
\hline
$2000 < Q^{2} < 8000$ & 38.96 & 3.8 & 5.9 & 2.2 / 0.3 & 37.02& 3.9 & 5.1& 1.2 / 1.0 \\
\hline
&\multicolumn{8}{c|}{ \rule{0mm}{2.5 mm} \raisebox{0.3mm}{$0.05 < x_{p} < 0.1$}}\\
\hline
$100 < Q^{2} < 175$ &   9.01& 1.8 & 3.2 & 1.6 / 1.1 & 8.81 & 1.8 & 3.3 & 1.5 / 0.8\\
\hline
$175 < Q^{2} < 250$ & 10.89 & 1.1 & 2.9 &  1.1 / 1.6 & 10.91 & 1.1 & 3.0 & 1.0 / 1.3 \\
\hline
$250 < Q^{2} < 450$ & 12.46 & 1.0 & 2.8 & 1.3 / 1.2 & 12.45 & 1.0 & 2.8 & 1.4 / 1.4 \\
\hline
$~450 < Q^{2} < 1000$ & 13.55 & 1.5 & 3.0 & 1.7 / 1.2 & 13.47 & 1.6 & 3.2 & 1.4 / 1.1 \\
\hline
$1000 < Q^{2} < 2000$ & 14.79 & 2.8 & 3.6 &  0.7 / 1.1 & 14.90 & 2.9 & 3.9 & 0.6 / 1.4 \\
\hline
$2000 < Q^{2} < 8000$ & 13.91 & 4.9 & 5.5 & 1.2 / 1.3 & 13.37 & 5.0 & 7.4 & 1.3 / 1.5 \\
\hline
&\multicolumn{8}{c|}{ \rule{0mm}{2.5 mm} \raisebox{0.3mm}{$0.1 < x_{p} < 0.2$}}\\
\hline
$100 < Q^{2} < 175$ & 4.69 & 1.7 & 3.2 &  1.1 / 1.1 & 4.68 & 1.7 & 3.3 & 1.2 / 0.8\\
\hline
$175 < Q^{2} < 250$ & 5.06 & 1.1& 2.8 & 1.7 / 1.5 & 5.08 & 1.1 & 2.9 & 1.6 / 2.0 \\
\hline
$250 < Q^{2} < 450$ & 5.37 & 1.1 & 3.0 & 1.2 / 1.6 & 5.17 & 1.1 & 3.1 & 1.5 / 2.0 \\
\hline
$~450 < Q^{2} < 1000$ & 5.44 & 1.7 & 3.0 & 1.2 / 1.0 & 5.39 & 1.8 & 3.1 & 1.2 / 1.4 \\
\hline
$1000 < Q^{2} < 2000$ & 5.55 & 3.3 & 3.8 & 1.0 / 1.4 & 5.08 & 3.5 & 6.0 & 1.0 / 1.2 \\
\hline
$2000 < Q^{2} < 8000$ & 5.16 & 5.8 & 6.7 & 1.6 / 2.3 & 4.67 & 6.2 & 7.3 & 1.1 / 2.3 \\
\hline
&\multicolumn{8}{c|}{ \rule{0mm}{2.5 mm} \raisebox{0.3mm}{$0.2 < x_{p} < 0.3$}}\\
\hline
$100 < Q^{2} < 175$ & 1.95 & 2.7 & 4.1 &  1.6 / 1.2 & 1.86 & 2.8 & 4.0 & 2.0 / 1.8 \\
\hline
$175 < Q^{2} < 250$ & 2.08 & 1.8 & 3.4 & 2.0 / 1.8 & 1.96 & 1.8 & 3.4 & 2.7 / 2.1 \\
\hline
$250 < Q^{2} < 450$ & 2.09 & 1.8 & 3.3 &  1.7 / 2.2 & 1.98 & 1.8 & 3.6 & 2.1 / 2.0 \\
\hline
$~450 < Q^{2} < 1000$ & 2.04 & 2.9 & 4.2 &  2.2 / 0.4 & 1.94 & 3.0 & 5.4 & 2.4 / 2.5 \\
\hline
$1000 < Q^{2} < 2000$ & 1.97 & 5.5 & 9.0 & 2.5 / 0.2 & 1.65 & 6.2 & 6.9 & 2.5 / 2.7 \\
\hline
$2000 < Q^{2} < 8000$ & 1.92 & 9.6 & 11.1 & 1.4 / 0.3 & 1.50 & 11.1 & 14.4 & 4.6 / 5.1 \\
\hline
&\multicolumn{8}{c|}{ \rule{0mm}{2.5 mm} \raisebox{0.3mm}{$0.3 < x_{p} < 0.4$}}\\
\hline
$100 < Q^{2} < 175$ & 0.910 & 4.0 & 5.0 &  1.1 / 1.9 & 0.832 & 4.1 & 6.1 & 1.5 / 2.2 \\
\hline
$175 < Q^{2} < 250$ & 0.925 & 2.7 & 4.0 & 2.8 / 3.7 & 0.892 & 2.7 & 4.0 & 3.1 / 2.6 \\
\hline
$250 < Q^{2} < 450$ & 0.955 & 2.6 & 4.5 & 3.9 / 2.3 & 0.856 & 2.8 & 4.0 & 3.0 / 2.9 \\
\hline
$~450 < Q^{2} < 1000$ & 0.943 & 4.2 & 5.0 & 2.9 / 3.0 & 0.713 & 4.9 & 5.9 & 1.8 / 1.8 \\
\hline
$1000 < Q^{2} < 2000$ & 0.807 & 8.8 &11.3 & 0.3 / 4.6 & 0.691 & 9.6 & 11.1 & 3.7 / 3.5 \\
\hline
$2000 < Q^{2} < 8000$ & 0.816 & 14.2 & 22.5 & 1.9 / 2.6 & 0.551 & 18.2 & 22.0 & 1.5 / 3.4 \\
\hline
&\multicolumn{8}{c|}{ \rule{0mm}{2.5 mm} \raisebox{0.3mm}{$0.4 < x_{p} < 0.5$}}\\
\hline
$100 < Q^{2} < 175$ & 0.453 & 5.7 & 8.3 & 2.9 / 1.8 & 0.395 & 5.9 & 7.1& 2.0 / 2.0 \\
\hline
$175 < Q^{2} < 250$ & 0.471 & 3.7 & 5.6 & 4.9 / 3.1 & 0.408 & 4.0 & 6.7& 3.1 / 3.8\\
\hline
$250 < Q^{2} < 450$ & 0.437 & 3.9 & 6.0 & 4.2 / 5.1 & 0.390 & 4.1 & 5.2 & 4.8 / 3.6 \\
\hline
$~450 < Q^{2} < 1000$ & 0.505 & 5.9 & 9.0 &  0.8 / 3.0 & 0.346 & 7.0 & 9.1 & 4.8 / 5.0 \\
\hline
$1000 < Q^{2} < 2000$ & 0.402 & 12.6 & 13.4 &  2.4 / 1.1 & 0.253 & 15.3 & 19.2 & 4.9 / 3.8 \\
\hline
$2000 < Q^{2} < 8000$ & 0.445 & 19.9 & 21.3 & 4.0 / 1.5 & 0.216 & 30.4 & 37.5 & 5.0 / 1.4 \\
\hline
&\multicolumn{8}{c|}{ \rule{0mm}{2.5 mm} \raisebox{0.3mm}{$0.5 < x_{p} < 0.7$}}\\
\hline
$100 < Q^{2} < 175$ & 0.171 & 6.4 & 11.1 &  3.3 / 2.3 & 0.164 & 6.7 & 7.9 & 3.5 / 1.3 \\
\hline
$175 < Q^{2} < 250$ & 0.191 & 4.2 & 6.3 & 3.5 / 6.3 & 0.146 & 4.6 & 5.9 & 6.6 / 6.9 \\
\hline
$250 < Q^{2} < 450$ & 0.178 & 4.4 & 6.6 & 4.8 / 4.9& 0.148 & 4.8 & 6.1 & 6.7 / 5.6 \\
\hline
$~450 < Q^{2} < 1000$ & 0.181 & 6.7 & 7.8 & 4.3 / 6.2 & 0.138 & 7.9 & 11.0 & 7.0 / 3.3\\
\hline
$1000 < Q^{2} < 2000$ & 0.143 & 14.2 & 15.9 & 4.3 / 4.7 & 0.106 & 16.6 & 17.5 & 3.7 / 1.9 \\
\hline
$2000 < Q^{2} < 8000$ & 0.194 & 21.7 & 24.2 & 9.0 / 3.6 & 0.104 & 29.1 & 29.5 & 0.8 / 5.5 \\
\hline
&\multicolumn{8}{c|}{ \rule{0mm}{2.5 mm} \raisebox{0.3mm}{$0.7 < x_{p} < 1.0$}}\\
\hline
$100 < Q^{2} < 175$ & 0.0324 & 12.2 & 13.1 & 3.8 / 5.1 & 0.0297 & 12.9 & 15.8 & 4.0 / 5.8\\
\hline
$175 < Q^{2} < 250$ & 0.0332 & 7.5 & 8.6 & 12.1 / 10.0 & 0.0240 & 9.0 & 10.2 & 13.4 / 14.5\\
\hline
$250 < Q^{2} < 450$ & 0.0337 & 8.2 & 8.9 & 9.6 / 7.5 & 0.0216 & 9.4 & 10.4 & 8.6 / 8.4 \\
\hline
$~450 < Q^{2} < 1000$ & 0.0329 &12.9 & 16.9 & 7.2 / 1.8 & 0.0200 & 15.8 & 19.9 & 4.6 / 12.7\\
\hline
$1000 < Q^{2} < 2000$ & 0.0274 & 26.4 & 51.0 & 1.0 / 5.0 & 0.0102 & 39.8 & 42.6 & 5.1 / 11.5 \\
\hline
$2000 < Q^{2} < 8000$ & 0.0181 & 43.6 & 48.2 & 13.7 / 8.3 & 0.0083 & 60.9 & 64.4 & 3.4 / 6.6 \\
\hline
\end{tabular}
\end{center}
\end{tiny}

\caption{\label{table:xpqsplit} The measured normalised distribution of the scaled momentum for positively, $D^{+}(x_{p},Q)$, and negatively, $D^{-}(x_{p},Q)$, charged particles, as a function of $Q^{2}$ for different $x_{p}$ intervals, shown with the statistical error ($\delta_{stat}$), the total error including statistical and systematic errors added in quadrature ($\delta_{tot}$ ), and the correlated error coming from the electron energy scale uncertainty ($\delta_{scale}$) which is shown as two numbers ($+/-$) and is not included in the total error. }
\end{table}%

\newpage

\begin{table}[htdp]
\renewcommand{\arraystretch}{1.2}
\begin{tiny}
\begin{center}
\begin{tabular}{|c||d|d|d|}
\hline
\multicolumn{1}{|c||}{$Q^{2}~[\GeV^{2}]$} & 
\multicolumn{1}{c|}{$A(x_{p},Q)$} & 
\multicolumn{1}{c|}{$\delta_{stat}$} & 
\multicolumn{1}{c|}{$\delta_{tot}$}  \\
\hline
\hline
&\multicolumn{3}{c|}{ \rule{0mm}{2.5 mm} \raisebox{0.3mm}{$0.0 < x_{p} < 0.02$}} \\
\hline
$100 < Q^{2} < 175$ & 0.007 & 0.036 & 0.045 \\
\hline
$175 < Q^{2} < 250$ & 0.005 & 0.019 & 0.032 \\
\hline
$250 < Q^{2} < 450$ & 0.010 & 0.015 & 0.030 \\
\hline
$~450 < Q^{2} < 1000$ & 0.021 & 0.016 & 0.032 \\
\hline
$1000 < Q^{2} < 2000$ & -0.019 & 0.022 & 0.034 \\
\hline
$2000 < Q^{2} < 8000$ & 0.028 & 0.027 & 0.039 \\
\hline
&\multicolumn{3}{c|}{ \rule{0mm}{2.5 mm} \raisebox{0.3mm}{$0.02 < x_{p} < 0.05$}}\\
\hline
$100 < Q^{2} < 175$ & -0.008 & 0.015 & 0.030\\
\hline
$175 < Q^{2} < 250$ & 0.013 &  0.009 & 0.027 \\
\hline
$250 < Q^{2} < 450$ & 0.003 & 0.008 & 0.028 \\
\hline
$~450 < Q^{2} < 1000$ & 0.012 & 0.010 & 0.028 \\
\hline
$1000 < Q^{2} < 2000$ & 0.001 & 0.017 & 0.033 \\
\hline
$2000 < Q^{2} < 8000$ & 0.017 & 0.027 & 0.039 \\
\hline
&\multicolumn{3}{c|}{ \rule{0mm}{2.5 mm} \raisebox{0.3mm}{$0.05 < x_{p} < 0.1$}}\\
\hline
$100 < Q^{2} < 175$ & 0.010 & 0.012 & 0.028 \\
\hline
$175 < Q^{2} < 250$ & 0.008 & 0.008 & 0.028 \\
\hline
$250 < Q^{2} < 450$ & 0.003 & 0.007 & 0.026 \\
\hline
$~450 < Q^{2} < 1000$ & 0.011 & 0.011 & 0.029 \\
\hline
$1000 < Q^{2} < 2000$ & 0.003 & 0.020 & 0.033 \\
\hline
$2000 < Q^{2} < 8000$ & 0.004 & 0.034 & 0.046 \\
\hline
&\multicolumn{3}{c|}{ \rule{0mm}{2.5 mm} \raisebox{0.3mm}{$0.1 < x_{p} < 0.2$}}\\
\hline
$100 < Q^{2} < 175$ & 0.008 & 0.012 & 0.029 \\
\hline
$175 < Q^{2} < 250$ & 0.001 & 0.008 & 0.027 \\
\hline
$250 < Q^{2} < 450$ & 0.020 & 0.008 & 0.027 \\
\hline
$~450 < Q^{2} < 1000$ & 0.003 & 0.012 & 0.028 \\
\hline
$1000 < Q^{2} < 2000$ & 0.025 & 0.024 & 0.042 \\
\hline
$2000 < Q^{2} < 8000$ & 0.056 & 0.041 & 0.051 \\
\hline
&\multicolumn{3}{c|}{ \rule{0mm}{2.5 mm} \raisebox{0.3mm}{$0.2 < x_{p} < 0.3$}}\\
\hline
$100 < Q^{2} < 175$ & 0.020 & 0.019 & 0.032 \\
\hline
$175 < Q^{2} < 250$ &  0.020  & 0.013 & 0.030 \\
\hline
$250 < Q^{2} < 450$ & 0.028 & 0.013 & 0.029 \\
\hline
$~450 < Q^{2} < 1000$ & 0.043 & 0.021 & 0.041 \\
\hline
$1000 < Q^{2} < 2000$ & 0.12 & 0.041 & 0.059 \\
\hline
$2000 < Q^{2} < 8000$ & 0.18 & 0.074 & 0.11 \\
\hline
&\multicolumn{3}{c|}{ \rule{0mm}{2.5 mm} \raisebox{0.3mm}{$0.3 < x_{p} < 0.4$}}\\
\hline
$100 < Q^{2} < 175$ & 0.058 &  0.029 & 0.041 \\
\hline
$175 < Q^{2} < 250$ & 0.018 & 0.019 & 0.032 \\
\hline
$250 < Q^{2} < 450$ & 0.049 & 0.019 & 0.034 \\
\hline
$~450 < Q^{2} < 1000$ & 0.15 & 0.032 & 0.045 \\
\hline
$1000 < Q^{2} < 2000$ & 0.047 & 0.062 & 0.083 \\
\hline
$2000 < Q^{2} < 8000$ & 0.32 & 0.12 & 0.18 \\
\hline
&\multicolumn{3}{c|}{ \rule{0mm}{2.5 mm} \raisebox{0.3mm}{$0.4 < x_{p} < 0.5$}}\\
\hline
$100 < Q^{2} < 175$ & 0.033 & 0.041 & 0.061 \\
\hline
$175 < Q^{2} < 250$ & 0.11 & 0.028 & 0.056 \\
\hline
$250 < Q^{2} < 450$ & 0.036 & 0.028 & 0.045 \\
\hline
$~450 < Q^{2} < 1000$ & 0.13 & 0.043 & 0.082 \\
\hline
$1000 < Q^{2} < 2000$ & 0.18 &  0.094 & 0.12 \\
\hline
$2000 < Q^{2} < 8000$ & 0.39 & 0.15 & 0.18 \\
\hline
&\multicolumn{3}{c|}{ \rule{0mm}{2.5 mm} \raisebox{0.3mm}{$0.5 < x_{p} < 0.7$}}\\
\hline
$100 < Q^{2} < 175$ & 0.070 & 0.047 & 0.074 \\
\hline
$175 < Q^{2} < 250$ & 0.12 & 0.031 & 0.048 \\
\hline
$250 < Q^{2} < 450$ & 0.081 & 0.032 & 0.047 \\
\hline
$~450 < Q^{2} < 1000$ & 0.095 & 0.048 & 0.068 \\
\hline
$1000 < Q^{2} < 2000$ & 0.19 & 0.11 & 0.12 \\
\hline
$2000 < Q^{2} < 8000$ & 0.30 & 0.17 &  0.19 \\
\hline
&\multicolumn{3}{c|}{ \rule{0mm}{2.5 mm} \raisebox{0.3mm}{$0.7 < x_{p} < 1.0$}}\\
\hline
$100 < Q^{2} < 175$ & 0.014 & 0.084 & 0.096 \\
\hline
$175 < Q^{2} < 250$ & 0.17 & 0.055 & 0.067 \\
\hline
$250 < Q^{2} < 450$ & 0.23 & 0.059 & 0.068 \\
\hline
$~450 < Q^{2} < 1000$ & 0.25 & 0.10   & 0.12  \\
\hline
$1000 < Q^{2} < 2000$ & 0.33 & 0.19 & 0.25  \\
\hline
$2000 < Q^{2} < 8000$ & 0.62 & 0.40 & 0.52 \\
\hline
\end{tabular}
\end{center}
\end{tiny}

\caption{\label{table:asymmetryq} The charge asymmetry, $A(x_{p},Q)$, as a function of $Q^{2}$ for different $x_{p}$ intervals shown with the statistical error ($\delta_{stat}$) and the total error including statistical and systematic errors added in quadrature ($\delta_{tot}$).}
\end{table}%

\newpage

\newpage

\newpage

\newpage

\end{document}